\documentclass{PoS}

\title{Lattice QCD study of baryon-baryon interactions in the (S,I)=(-2,0) system using the coupled-channel formalism}

\ShortTitle{Lattice QCD study of baryon-baryon interactions in the (S,I)=(-2,0) system }

\author{\speaker{K. Sasaki} \\
        Graduate School of Pure and Applied Sciences, University of Tsukuba,\\
        Tsukuba, Ibaraki 305-8577, Japan\\
        E-mail: \email{kenjis@het.ph.tsukuba.ac.jp}}

\author{for HAL QCD collaboration \vspace*{-0.2cm} \\ 
\begin{center} \includegraphics[width=.35\textwidth]{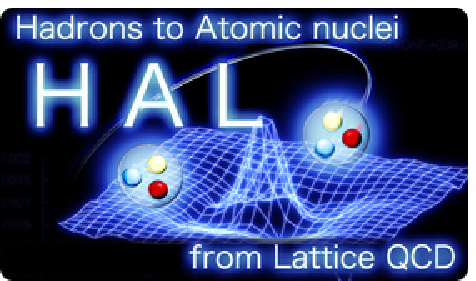} \end{center}}

\abstract{We investigate baryon-baryon interactions with strangeness $S=-2$ and isospin $I=0$ system from Lattice QCD.
In order to solve this system, we prepare three types of baryon-baryon operators ($\Lambda-\Lambda$, $N-\Xi$ and $\Sigma-\Sigma$) for the sink and construct three source operators diagonalizing the $3\times3$ correlation matrix.
Combining of the prepared sink operators with the diagonalized source operators, we obtain nine effective Nambu-Bethe-Salpeter (NBS) wave functions.
The $3\times3$ potential matrix is calculated by solving the coupled-channel Schr\"odinger equation.
The flavor $SU(3)$ breaking effects of the potential matrix are also discussed by comparing with the results of the $SU(3)$ limit calculation.
Our numerical results are obtained from three sets of $2+1$ flavor QCD gauge configurations provided by the CP-PACS/JLQCD Collaborations.}

\FullConference{The XXVIII International Symposium on Lattice Field Theory, Lattice2010\\
		June 14-19, 2010\\
		Villasimius, Italy}

\begin{document}

\section{Introduction}
Completion of the knowledge of the generalized nuclear force, which includes not only the nucleon-nucleon ($NN$) interaction but also hyperon-nucleon ($YN$) and hyperon-hyperon ($YY$) interactions, brought the deeper understanding of atomic nuclei, structure of neutron stars and supernova explosions.
However it is hard to know the properties of the $YN$ and $YY$ interactions because the $YN$ and $YY$ scattering data in free-space are scarce.

The satisfactory theoretical description of the generalized nuclear force has not yet been presented because experimental studies of the $YN$ and $YY$ scatterings are extremely limited due to the lack of the hyperon beam.

Recently a method to extract the $NN$ potential through the Nambu-Bethe-Salpeter (NBS) wave function from lattice QCD simulations has been proposed in~\cite{Ishii:2006ec}.
The obtained potential is found to have desirable features, such as attractive well at long and medium distances, and the central repulsive core at short distance~\cite{Ishii:2006ec,Aoki:2009ji}. 
The method has been applied to the $YN$ systems such as  $\Lambda N$ and $\Xi N$ systems~\cite{Nemura:2008sp,Nemura:2009kc,Nemura:2010nh}.
Further investigation of baryon-baryon (B-B) interaction have been done in the flavor $SU(3)$ symmetric world, and the flavor-spin dependence of S-wave B-B interactions are revealed in Ref.~\cite{Inoue:2010hs}.

In this work, we focus on the $S=-2$, $I=0$ B-B system to seek the $\Lambda \Lambda$ interaction and to see the flavor $SU(3)$ breaking effects of the B-B interaction from Lattice QCD simulation.
The interesting point of this system is that the flavor singlet state is involved.
The flavor singlet state is known to generate the strong attraction and free from the Pauli blocking at the short range region from the view of quark degrees of freedom~\cite{Jaffe:1976yi,Oka:1983ku}.
For this reason we would like to know how its effects appear in the real world.

\section{Formulation}
The $(S,I)=(-2,0)$ B-B state consists of the $\Lambda \Lambda$, $N \Xi$ and $\Sigma \Sigma$ components in terms of low-lying baryons.
Mass differences of these components are quite small, and it causes a mixture of NBS wave function from excited states.
In sucn situation the source operator should be optimized to extract the energy eigen states through the variational method~\cite{Michael:1985ne,Luscher:1990ck}.

Using diagonalized source operators, we obtain the NBS wave function of $\Lambda \Lambda$, $N \Xi$ and $\Sigma \Sigma$ components for each energy.
The transition potential matrix of 3-states coupled channel equation can be acquired
\begin{eqnarray}
{V^{i}}_j \equiv
 \left( \begin{array}{ccc}
 {V^{\Lambda \Lambda}}_{\Lambda \Lambda} & {V^{\Lambda \Lambda}}_{N \Xi} & {V^{\Lambda \Lambda}}_{\Sigma \Sigma} \\
 {V^{N \Xi}}_{\Lambda \Lambda} & {V^{N \Xi}}_{N \Xi} & {V^{N \Xi}}_{\Sigma \Sigma} \\
 {V^{\Sigma \Sigma}}_{\Lambda \Lambda} & {V^{\Sigma \Sigma}}_{N \Xi} & {V^{\Sigma \Sigma}}_{\Sigma \Sigma} \\
 \end{array} \right).
\end{eqnarray}
This is an application of an extended HAL QCD method formulated in Ref.~\cite{Ishii:2010lt}.

\subsection{Diagonalization of source operators}
We prepare three wall-source operators at $t=t_0$ as
\begin{eqnarray}
{\cal{J}}^{\Lambda \Lambda}(t_0) = P^S_{\alpha \beta} [\Lambda_\alpha \Lambda_\beta](t_0),
~
{\cal{J}}^{N \Xi}(t_0) = P^S_{\alpha \beta} [N_\alpha\Xi_\beta](t_0) 
~ {\rm{and}} ~
{\cal{J}}^{\Sigma \Sigma}(t_0) = P^S_{\alpha \beta} [\Sigma_\alpha \Sigma_\beta](t_0) 
\end{eqnarray}
where the square bracket stands for the $I=0$ combination of two baryons. 
The $P^S$ is the projection operator of spin singlet state. 
We use the same definition for the interpolating field operators as given in Ref.~\cite{Inoue:2010hs}.

In order to create the energy eigen states, we make a linear combination of three source operators as
\begin{eqnarray}
 {\cal{I}}_\alpha(t) = v^{\Lambda \Lambda}_\alpha {\cal{J}}^{\Lambda \Lambda}(t)
 + v^{N \Xi}_\alpha {\cal{J}}^{N \Xi}(t)
 + v^{\Sigma \Sigma}_\alpha {\cal{J}}^{\Sigma \Sigma}(t).
\end{eqnarray}
Then we define the effective mass of the operator ${\cal{I}}$,
\begin{eqnarray}
 m(\Delta t) \equiv - \frac{1}{\Delta t} \ln \left[ \frac{\langle  {\cal{I}}(t+\Delta t) \bar{\cal{I}}(0) \rangle}{\langle  {\cal{I}}(t) \bar{\cal{I}}(0) \rangle} \right]
\end{eqnarray}
where $\Delta t$ is usually taken to be $1$.
The stationary condition of the effective mass against the coefficients $v$ leads 
\begin{eqnarray}
 {\bf{C}}(t+\Delta t) {\bf{v}}                
=
e^{-m(\Delta t) \cdot \Delta t} {\bf{C}}(t) {\bf{v}} 
\end{eqnarray}
with the correlation matrix defined as
\begin{eqnarray}
 C^{IJ}(t) \equiv \langle {\cal{J}}^I(t) \bar{\cal{J}}^J(0) \rangle
 \hspace*{1em} {\small{(I,J=\Lambda \Lambda, N \Xi, \Sigma \Sigma)}}.
\end{eqnarray}
The optimized source operators ${\cal{I}}_1$, ${\cal{I}}_2$ and ${\cal{I}}_3$ which strongly couples to the ground, 1st excited and 2nd excited state respectively are obtained by solving the generalized eigen-problem.

\subsection{Coupled channel Schr\"odinger equarion}
The equal-time NBS wave function $\psi^{B_1B_2}(\vec{r},E)$ for an energy eigen state with $E$ is extracted from the four point function,
\begin{equation}
 W^{B_1B_2}(t-t_{0},\vec r)
= \sum_{\vec x} \langle 0 \mid B_1(t,\vec x+\vec r)\,B_2(t,\vec x)\,
                       {\bar {\cal{I}}_E}(t_0) \mid 0 \rangle ~,
\end{equation}
where ${\cal{I}}_E$ is diagonalized wall-source operator which is expected to separate the energy eigen state from the other one.

General form of the Schr\"odinger equation for the full wave function $\psi(\vec r, E)$ with energy $E$ is given in the form with the free Hamiltonian $H_0$ as
\begin{equation}
   \left[ H_0 - E \right] \psi(\vec r, E) 
 = \int\!\!d^3\vec r' \, U(\vec r, \vec r') \, \psi(\vec r', E)
\end{equation}
where $U(\vec r, \vec r')$ is an energy independent but non-local potential.
 At low energies, it can be expanded by the local velocity $\vec v = \vec{\nabla}/\mu$ as   
  $U(\vec r,\vec r') = V(\vec r, \vec{\nabla})\delta(\vec r-\vec r') 
  = (V_{LO}+ V_{NLO} + V_{NNLO} + \cdots )\delta(\vec r-\vec r') $,
  where $N^nLO$ term is of $O(v^n)$.
It is shown in Ref.~\cite{Murano:2010tc} that the LO potential $V(\vec r)$ defined above
is a good approximation of $U(\vec r,\vec r')$ at low kinetic energies
 between $T \sim 0$ MeV  and $T \sim 45$ MeV for the $NN$ system in quenched QCD.

Taking the leading order of velocity expansion of non-local interaction kernel and projection operators into the $\Lambda \Lambda$, $N \Xi$ and $\Sigma \Sigma$ components, we have the coupled channel effective Schr\"odinger equation as 
\begin{eqnarray}
  \label{EQ:CoupledSE}
\begin{array}{ll}
\begin{array}{l}
\left( \displaystyle{\frac{\nabla^2}{ 2 \mu_{\Lambda \Lambda}} + \frac{p^2_i}{ 2 \mu_{\Lambda \Lambda}}} \right) \psi^{\Lambda \Lambda}(\vec{r}, E_i)
  = \displaystyle\sum_\gamma {V^{\Lambda \Lambda}}_{\gamma}(\vec{r}) \psi^{\gamma}(\vec{r}, E_i)
\\
\left( \displaystyle{\frac{\nabla^2}{ 2 \mu_{N \Xi}} + \frac{q^2_i}{ 2 \mu_{N \Xi}}} \right) \psi^{N \Xi}(\vec{r}, E_i)
  = \displaystyle\sum_\gamma {V^{N \Xi}}_{\gamma}(\vec{r}) \psi^{\gamma}(\vec{r}, E_i)
\\
\left( \displaystyle{\frac{\nabla^2}{ 2 \mu_{\Sigma \Sigma}} + \frac{k^2_i}{ 2 \mu_{\Sigma \Sigma}}} \right) \psi^{\Sigma \Sigma}(\vec{r}, E_i)
  = \displaystyle\sum_\gamma {V^{\Sigma \Sigma}}_{\gamma}(\vec{r}) \psi^{\gamma}(\vec{r}, E_i)
\end{array} &
\begin{array}{l}
\hspace*{1em} {\small{(i=0,1,2)}} \\
\hspace*{1em} {\small{(\gamma=\Lambda \Lambda, N \Xi, \Sigma \Sigma)}} 
\end{array}
\end{array}
\end{eqnarray}
where asymptotic momenta are related to energies of state as
\begin{eqnarray}
E_i = 2 \sqrt{m_\Lambda^2 + p_i^2} = \sqrt{m_N^2+q_i^2} + \sqrt{m_\Xi^2+q_i^2} = 2 \sqrt{m_\Sigma^2 + k_i^2}.
\end{eqnarray}
Solving the eq.~(\ref{EQ:CoupledSE}) for the transition potentials, we have
\begin{eqnarray}
\left( \begin{array}{c}
 {V^{\Lambda \Lambda}}_{\Lambda \Lambda}(\vec{r}) \\
 {V^{\Lambda \Lambda}}_{N \Xi}(\vec{r}) \\
 {V^{\Lambda \Lambda}}_{\Sigma \Sigma}(\vec{r})
\end{array} \right)
=
\frac{1}{2\mu_{\Lambda \Lambda}}
\left( \begin{array}{ccc} 
\psi^{\Lambda \Lambda}(\vec{r},E_0)  & \psi^{N \Xi}(\vec{r},E_0)  & \psi^{\Sigma \Sigma}(\vec{r},E_0)  \\ 
\psi^{\Lambda \Lambda}(\vec{r},E_1)  & \psi^{N \Xi}(\vec{r},E_1)  & \psi^{\Sigma \Sigma}(\vec{r},E_1)  \\
\psi^{\Lambda \Lambda}(\vec{r},E_2) & \psi^{N \Xi}(\vec{r},E_2)  & \psi^{\Sigma \Sigma}(\vec{r},E_2)  \\ 
\end{array} \right)^{-1}
\left( \begin{array}{c}
 \left( \nabla^2 + p^2_0  \right) \Psi^{\Lambda \Lambda}(\vec{r},E_0) \\
 \left( \nabla^2 + p^2_1  \right) \Psi^{\Lambda \Lambda}(\vec{r},E_1) \\
 \left( \nabla^2 + p^2_2  \right) \Psi^{\Lambda \Lambda}(\vec{r},E_2)
\end{array} \right).
\nonumber \\
\end{eqnarray} 
The other transition potentials are obtained in a similar manner.

\section{Numerical setup}
\begin{table}
\begin{center}
\caption{Lattice parameters and hadron masses in unit of [MeV] are listed.}
\label{TAB:HadronM}
 \begin{tabular}{cccccc}
\hline \hline
 \multicolumn{6}{c}{Lattice parameters} \\
 $\beta$ & $\kappa_s$ & $c_{SW}$ & lattice size & $a$ [fm] & $L$ [fm] \\
\hline
 1.83 & 0.13710 & 1.7610 & $16^3 \times 32$ & $0.121$ & $1.93$ \\
\hline \hline
 \end{tabular} \\[2mm]
  \begin{tabular}{ccccccccc}
  \hline \hline
   & $N_{conf}$ & $\kappa_{ud}$ & $m_\pi$ & $m_K$ & $m_N$ & $m_\Lambda$ & $m_\Sigma$ & $m_\Xi$ \\
  \hline
  Set 1 & $700$ & $0.13760$ & $875(1)$ & $916(1)$ & $1806(3)$ & $1835(3)$ & $1841(3)$ & $1867(2)$ \\
  Set 2 & $800$ & $0.13800$ & $749(1)$ & $828(1)$ & $1616(3)$ & $1671(2)$ & $1685(2)$ & $1734(2)$ \\
  Set 3 & $800$ & $0.13825$ & $661(1)$ & $768(1)$ & $1482(3)$ & $1557(3)$ & $1576(3)$ & $1640(3)$ \\
  \hline \hline
  \end{tabular}
 \end{center}
\end{table}
In this calculation we employ $2+1$-flavor full QCD gauge configurations of Japan Lattice Data Grid(JLDG)/International Lattice Data Grid(ILDG)~\cite{JLDGILDG}.
They are generated by the CP-PACS and JLQCD Collaborations with a renormalization-group improved gauge action and a non-perturbatively $O(a)$ improved clover quark action at $\beta = 6/g^2=1.83$, corresponding to lattice spacings of $a = 0.1209~{\rm{ fm}}$~\cite{Ishikawa:2007nn}.
We choose three ensembles of the $L^3 \times T = 16^3 \times 32$ lattice which means the spatial volume of about $(2.0~{\rm{fm}})^3$. 
These ensembles are called as Set 1, 2, and 3 corresponding to the hopping parameter for light quarks $\kappa_{u,d} = 0.13760, 0.13800~{\rm{and}}~0.13825$, respectively, keeping $\kappa_{s} = 0.13710$ fixed for the $s$-quark in all ensemble. 
Quark propagators are calculated from the spatial wall source at $t_0$ with the Dirichlet boundary condition in temporal direction at $t - t_0 = 16$. 
The wall source is placed at 16 different time slices on each of different gauge configuration ensembles, in order to enhance the signals, together with the average over forward and backward propagations in time. 
The average over discrete rotations of the cubic group is taken for the sink operator, in order to obtain the relative S-wave in the B-B wave function. 
The numerical computation is carried out at KEK supercomputer system, Blue Gene/L.
The hadron masses are shown in Table~\ref{TAB:HadronM}.

\section{Result}
\begin{figure}
 \begin{tabular}{ccc}
 Ground state & 1st excited state & 2nd excited state \\
 \scalebox{0.33}[0.2]{\includegraphics{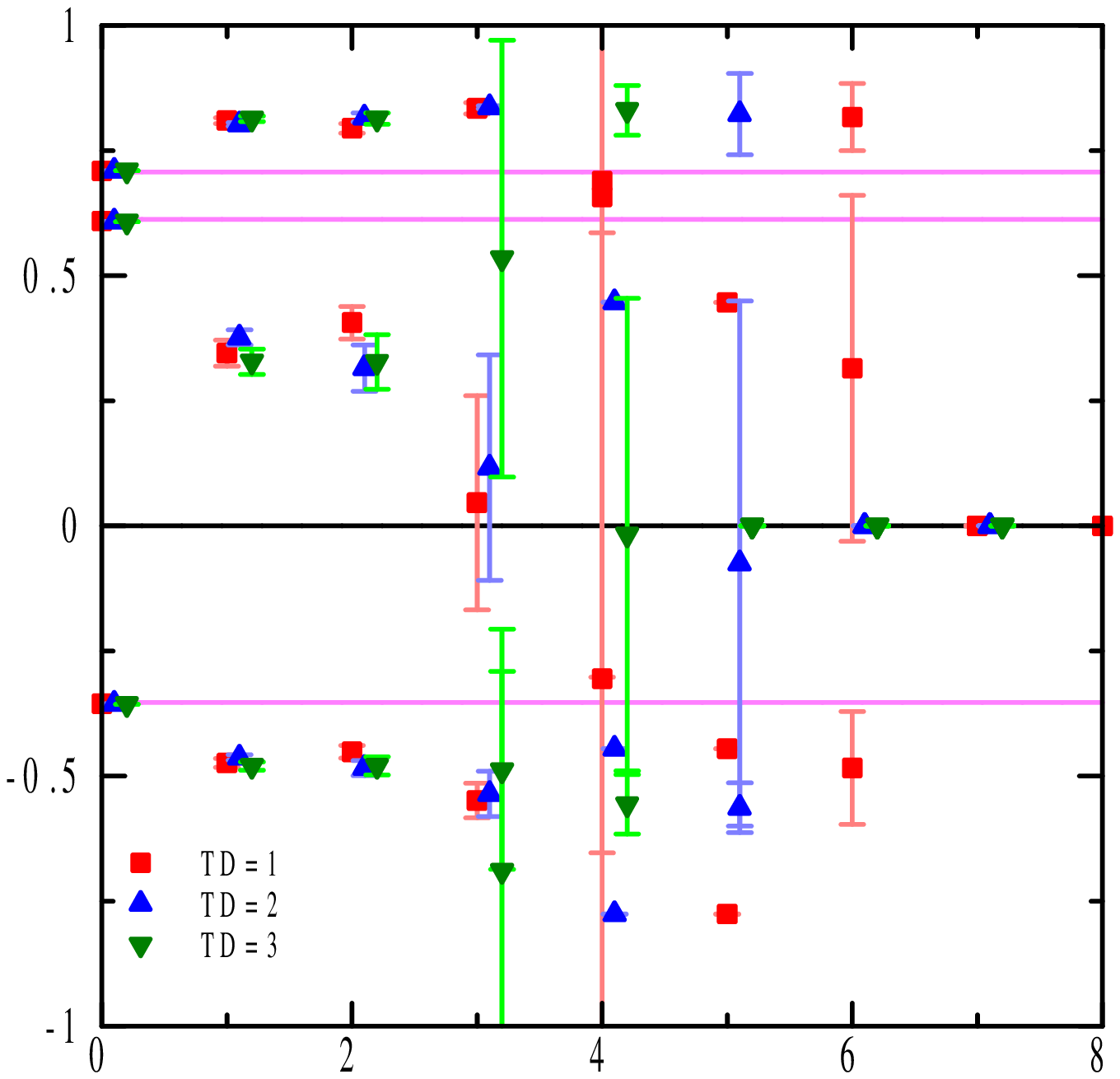}} &
 \scalebox{0.33}[0.2]{\includegraphics{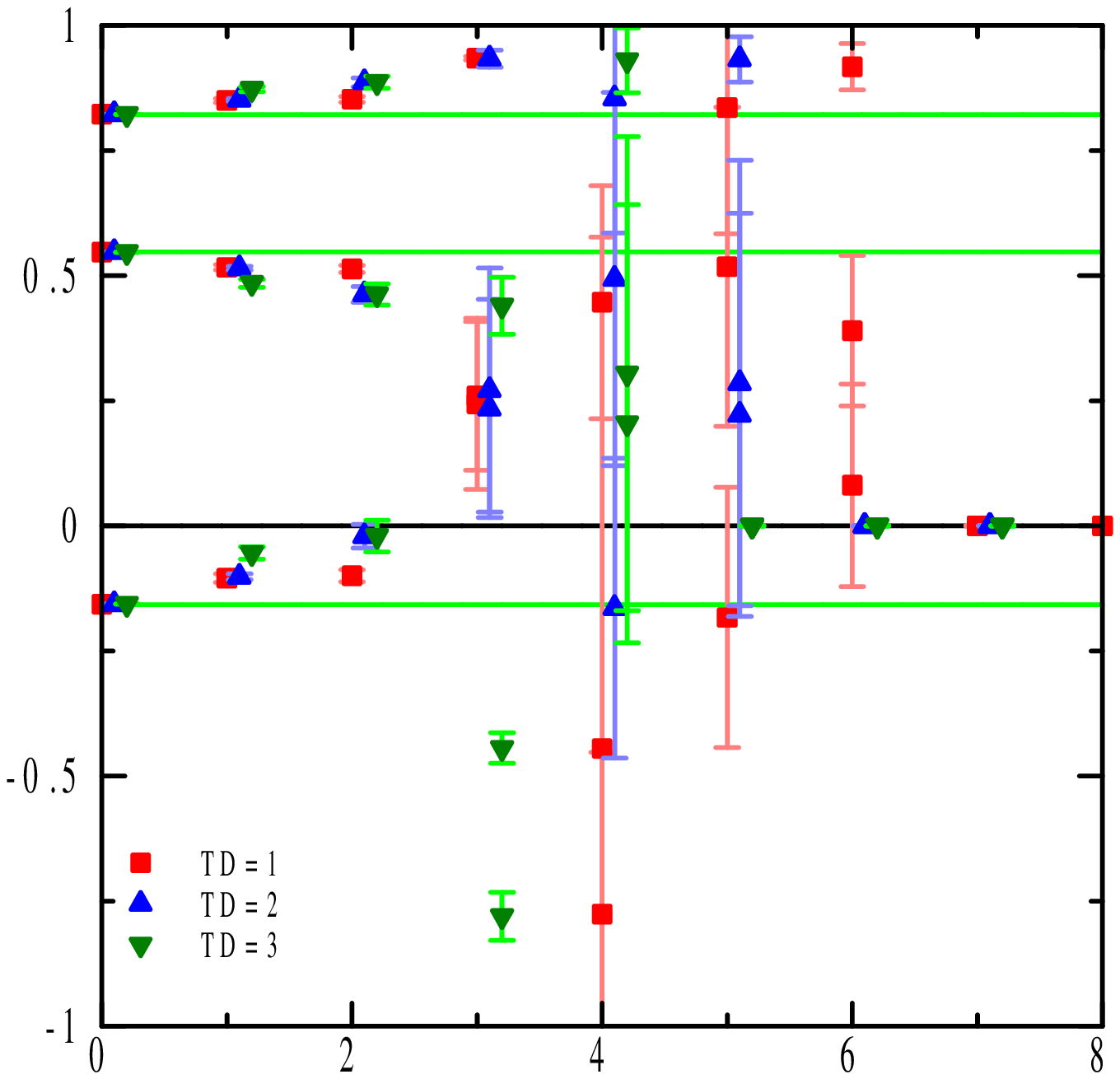}} &
 \scalebox{0.33}[0.2]{\includegraphics{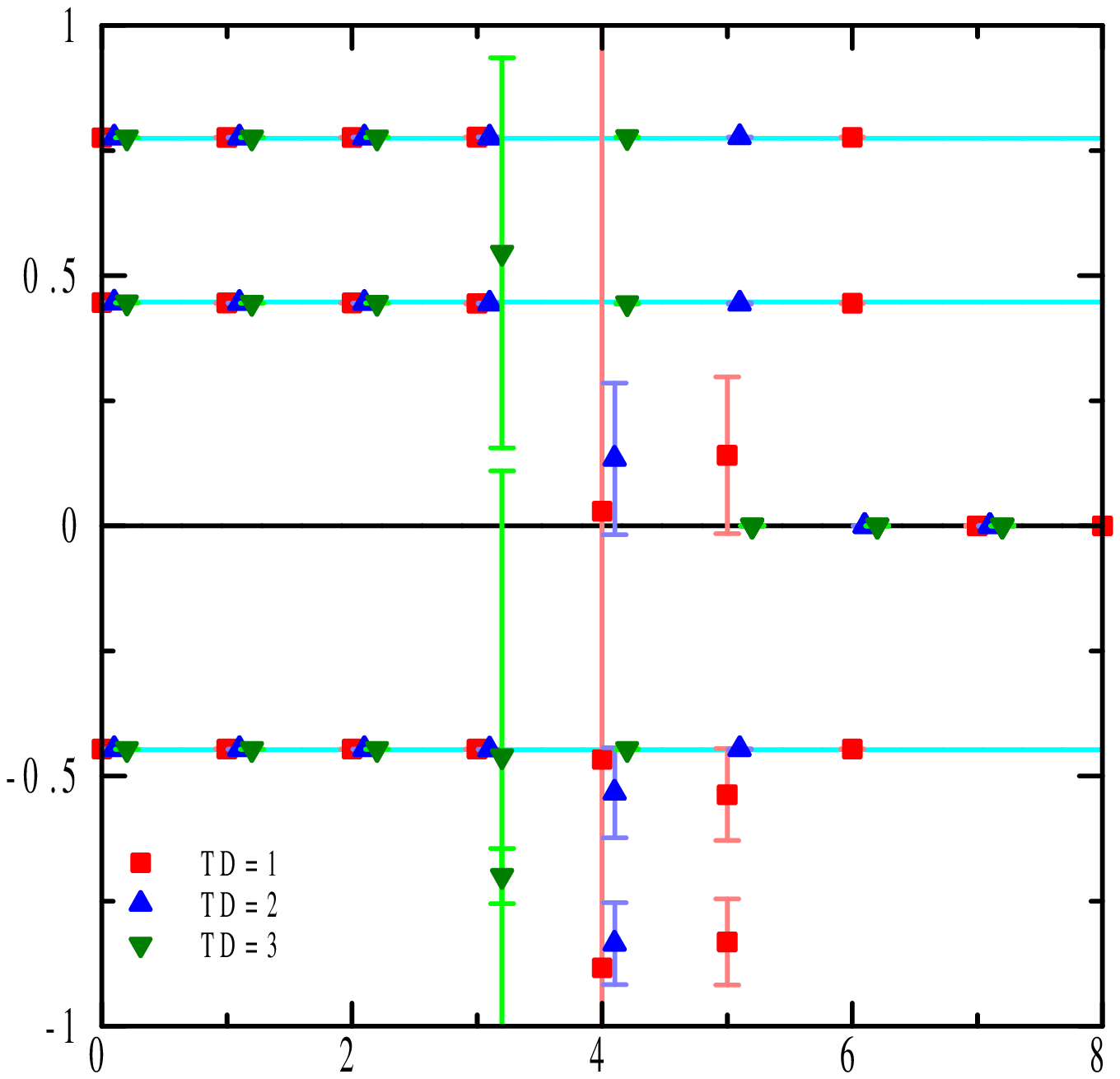}} \\[-10pt]
 {\tiny{t}} & {\tiny{t}} & {\tiny{t}}
 \end{tabular}
\caption{Eigen vectors of configuration Set 1. Three panels shows the eigen vector of ground, 1st excited and 2nd excited state from the left, respectively. Solid lines stand for the $SU(3)$ Clebsh-Gordan coefficients. The "TD" in figures stands for the $\Delta t$ in the text.}
\label{FIG:EVSet1}
\end{figure}

Figure~\ref{FIG:EVSet1} shows the eigen vectors of configuration Set 1.
From this figure it is difficult to adopt the values in the deep time slice region due to the large noise of the wall-to-wall correlation matrix though they are well determined in the shallow time slice region.
In this calculation we adopt a set of eigen vectors at the shallow time slice but we expect that they could be extrapolate to the deep time slice region around $t=9$ or $10$ because the eigen vectors are not change so much against $t$.

\begin{figure}
 \begin{tabular}{c}
 \scalebox{0.47}[0.34]{\includegraphics{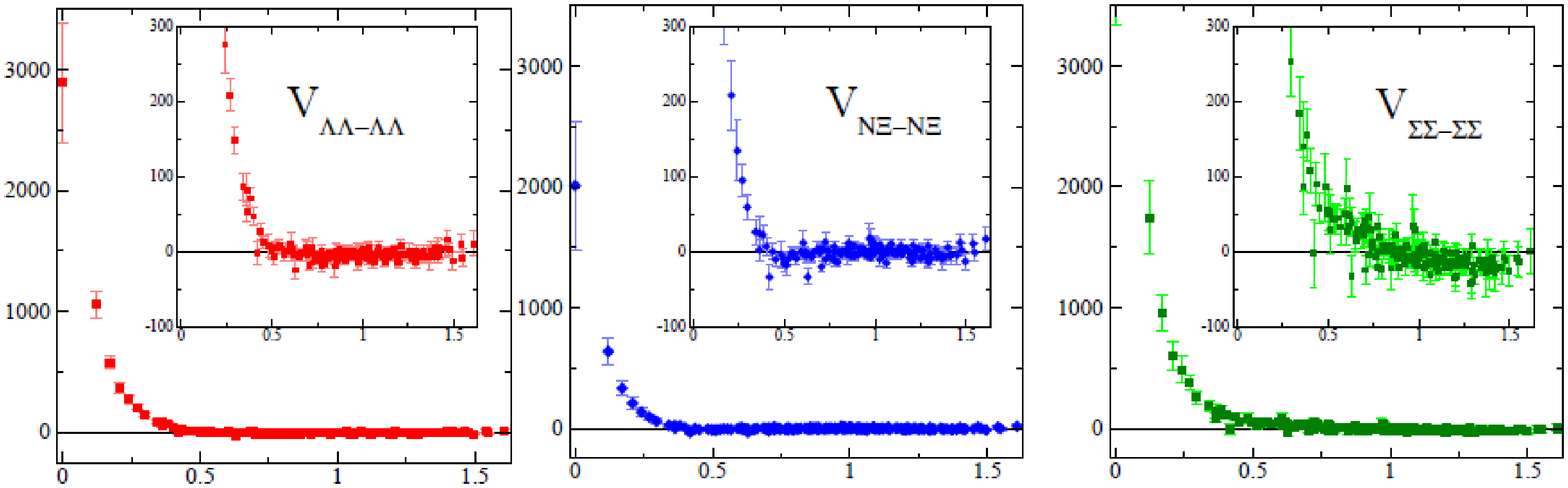}} \\
 \scalebox{0.47}[0.34]{\includegraphics{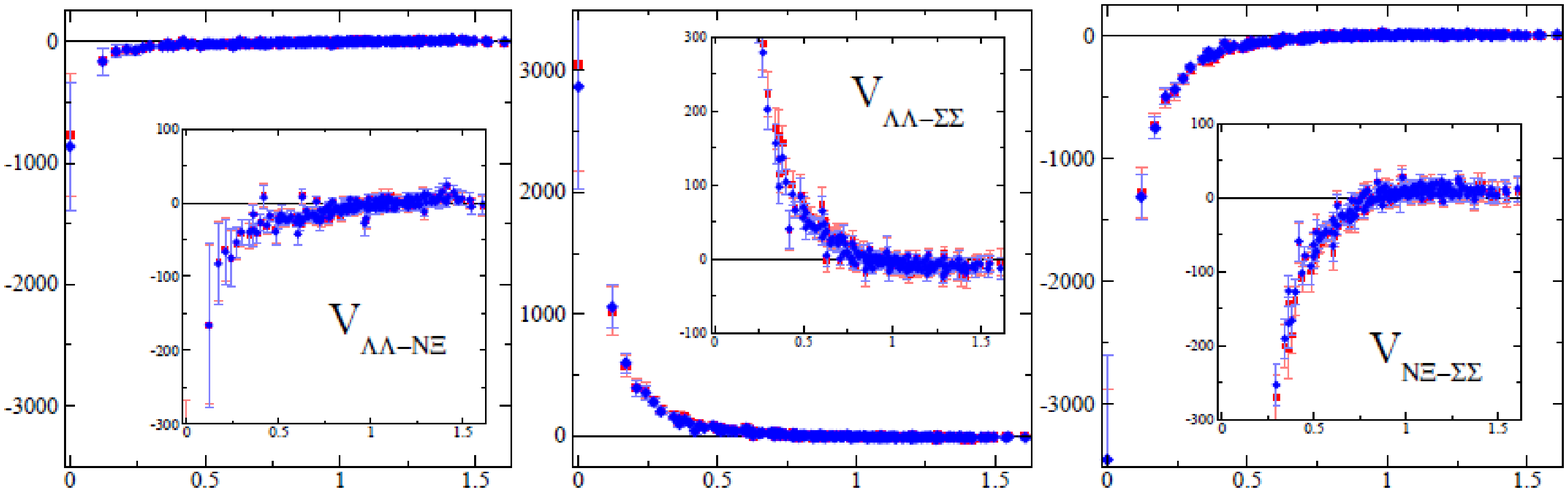}} 
 \end{tabular}
\caption{The potential matrix in Set 1 using the NBS wave function at $t=9$ is plotted. The vertical axis is for the potential strength in unit of [MeV], and the horizontal for the relative distance between two baryons in unit of [fm]. }
\label{FIG:PTpbSet1}
\end{figure}

The potential matrix ${V^I}_J$ in Set 1 is shown in Figure~\ref{FIG:PTpbSet1} using the NBS wave function at $t-t_0=9$.
We can see the flavor dependence of the height of repulsive core at short distance region. 
The $\Sigma \Sigma$ potential has the strongest repulsive core of these three channels.
In off-diagonal parts of potential matrix, we find that the strengths of ${V^{\Lambda \Lambda}}_{\Sigma \Sigma}$ and ${V^{N \Xi}}_{\Sigma \Sigma}$ are similar but it of ${V^{\Lambda \Lambda}}_{N \Xi}$ is much weaker than the others.

In order to compare the results of potential matrix calculated in three configuration sets, we transform the potentials from the particle basis to the $SU(3)$ irreducible representation (IR) basis as
\begin{eqnarray}
 V^{IR} = U V U^t = 
 \left( \begin{array}{ccc}
 {V^{1}} & {V^{1}}_{8} & {V^{1}}_{27} \\
 {V^{8}}_{1} & {V^{8}} & {V^{8}}_{27} \\
 {V^{27}}_{1} & {V^{27}}_{8} & {V^{27}} \\
 \end{array} \right)
\end{eqnarray}
where $U$ is an unitary transformation matrix whose explicit form is given in Appendix B in Ref.~\cite{Inoue:2010hs}.
The potential matrix in the IR basis is convenient and a good measure of the $SU(3)$ breaking effect by comparing three configuration sets because it should be diagonal in the $SU(3)$ symmetric limit.

\begin{figure}
 \begin{center}
  \scalebox{0.47}[0.34]{\includegraphics{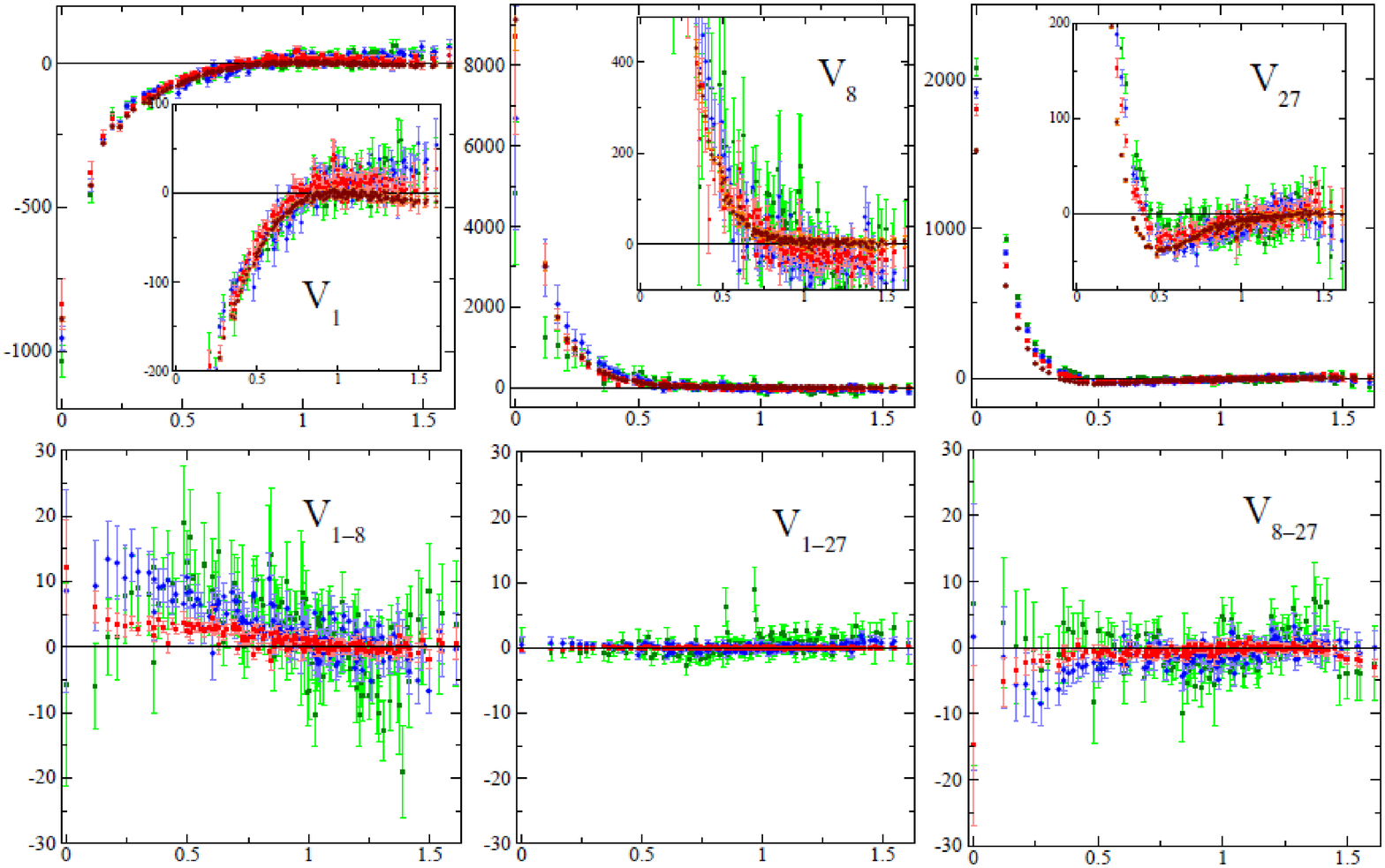}}
 \end{center}
\caption{Transition potentials in the $SU(3)$ IR basis. Red, blue and green symbol respectively correspond to the result of Set1, Set2 and Set3. The results of the $SU(3)$ symmetric configuration with $\kappa_{uds}=0.13710$ are also plotted with brown symbol~\cite{Inoue:2010hs}.}
\label{FIG:potall}
\end{figure}

In Figure~\ref{FIG:potall} we compare the results of the potential matrix in the IR basis calculated in different configuration sets.
We also plot the results of the potential matrix in the $SU(3)$ symmetric limit on top of them.
Diagonal parts of the potential matrix can be compared with the $SU(3)$ symmetric result.
We found the growth of repulstive core in the $V^{27}$ potential with decreasing the light quark mass.
The $V_{1-27}$ and $V_{8-27}$ transition potentials are consistent with zero within error bar.
On the other hand, it is noteworthy that the $V_{1-8}$ transition potential which is not allowed in the $SU(3)$ symmetric world is appearing.

\section{Conclusion}
We have investigated the $(S,I) = (-2,0)$ B-B state, which is known as the $\Lambda \Lambda$, $N \Xi$ and $\Sigma \Sigma$ coupled state, from lattice QCD.
In order to deal with this complicated system, we have devised an extention of HAL QCD method \cite{Ishii:2010lt} of construct the interaction potential to a coupled channel system.
First, we have diagonalized source operators which enable us to obtain the NBS wave functions with each energy.
Second, we have put all component of the NBS wave function in the coupled channel equation for the considering system.
Then we could obtain the desired potential matrix.
Although there have been the challanging problem to determine the eigen vector of source operators, we have confirmed that this technique is working well.

We have searched the $SU(3)$ breaking effects of the $(S,I) = (-2,0)$ BB state by changing the light-flavored quark mass.
We have found a small transition potential in $V_{1-8}$ in terms of the $SU(3)$ IR basis.
Such transition can not be allowed in the $SU(3)$ symmetric world.

This method could greatly assist us to complete the knowledge of not only the generalized nuclear force but also the interaction of hadrons including mesons, baryons and quarks.

\section*{Acknowledgement}
We are grateful for authors and maintainers of CPS++~\cite{CPS}, and which a modified version is used for simulation done in this work. 
We wish to acknowledge both ILDG/JLDG and the collaboration providing the dataset~\cite{JLDGILDG}.
This work was supported by the Large Scale Simulation Program No.0923(FY2009) of High Energy Accelerator Research Organization (KEK), Grant-in-Aid of the Ministry of Education, Science and Technology, Sports and Culture (Nos. 20340047, 22540268, 19540261) and the Grant-in-Aid for Scientific Research on Innovative Areas (No. 2004:20105001,20105003).

\end{document}